\documentstyle[psfig,amssym]{l-aa}

\begin{document}
 
\thesaurus{12 (12.03.1; 12.03.3; 12.03.4; 12.12.1)}
 
\title{Gravitationally Lensed Sub-mm Sources towards Galaxy Clusters}
\author{Asantha R. Cooray}
\institute{Department of Astronomy and Astrophysics, University of Chicago, 
Chicago IL 60637, USA. E-mail: asante@hyde.uchicago.edu}
\date{Received: August 19, 1998; accepted }
\maketitle


\begin{abstract} 

Recent observations of galaxy clusters with the SCUBA instrument 
on the James Clarke Maxwell Telescope have revealed a sample
 of gravitationally lensed sources at sub-mm wavelengths. 
We extend our previous calculation on the expected number of lensed optical
arcs and radio sources to study the sub-mm lensed source statistics due to
foreground massive clusters.
For a flat cosmology with $\Omega_m=0.4$ and $\Omega_\Lambda =0.6$,
we predict $\sim 1.5 \times 10^{4}$ 
lensed sub-mm sources with flux densities greater than
 $\sim$  4  mJy at 850 $\mu$m, and with amplifications due to 
lensing greater than 2,
towards galaxy clusters with X-ray luminosities
greater than $8 \times 10^{44}\, h^{-2}\, {\rm ergs\, s^{-1}}$ (0.3 to 3.5
keV band).  We compare our predictions with the observations
 from the SCUBA instrument, and briefly consider the
possibility of using the South Pole 10-m sub-mm telescope and the
Planck surveyor to identify lensed sub-mm sources.
A catalog of around 100 gravitationally lensed sources 
at 353 GHz may be a useful by-product of Planck.

\end{abstract}


\section{Introduction}

It is now well known that gravitational lensing statistics is a useful
probe of the geometry of the universe, especially for the determination
of the cosmological constant.  In a recent paper,
 we calculated the expected number
of gravitationally lensed optical arcs and radio sources (Cooray 1998; hereafter C98)
on the sky due to foreground galaxy clusters as a function of the
cosmological parameters. The expected
number of lensed sources was calculated based on  the redshift distribution
of the Hubble Deep Field (HDF; Williams et al. 1996) galaxies,
with an extrapolation to the whole sky. 
Here, we extend our calculations to estimate the number of
expected sub-mm sources on the whole sky due to foreground
clusters. Our calculation is prompted by recent observational
results from the new Sub-millimeter Common-User 
Bolometer Array (SCUBA; see, e.g., Cunningham et al. 1994)
on the James Clerk Maxwell Telescope, where a sample of gravitationally
lensed sub-mm sources has now been observed by Smail et al. (1997 \& 1998).

The gravitational lensing of background sub-mm sources due to
foreground galaxy clusters was first studied by Blain (1997), using
a model of a lensing cluster with predicted source counts  for background
sources. Blain (1997) showed that the surface and flux densities
of lensed sources exceed those of galaxies within the lensing cluster, and
their values. This behavior, primarily due to the slope of the
 source counts and the fact that the 
distance sources are intrinsically brighter
at sub-mm wavelengths, has now allowed the observation of moderate to
high redshift dusty star forming galaxies, which are
 amplified through the cluster potentials
(e.g., Ivison et al. 1998).
Even though lensing of sub-mm sources have been studied in literature,
no clear prediction has been made on the total number of sources
 lensed due to foreground clusters. Also,
past calculations have relied mostly  on models of background source
counts, which were based on different evolutionary scenarios
for star forming galaxies. Even though background source evolution
has been considered in these calculations,
 no attempt has  been made to consider the redshift evolution of
foreground lensing clusters.

 Thus, past calculations are limited if one were to
consider the possibility of using lensed sub-mm source statistics towards
clusters to constrain the cosmological parameters. We consider an
alternative approach, primarily based on the observational data
to study the number statistics of lensed sub-mm sources on the whole
sky due to galaxy clusters. We calculate
 the 
foreground lensing galaxy cluster number density and redshift  distribution
using a Press-Schechter (PS; Press \& Schechter 1974) analysis for
different cosmological models
and describe the background sub-mm sources using
 observational data towards the HDF
at 850 $\mu$m (Hughes et al. 1998).

In \S~2 we briefly describe our calculation and its inputs.  
In \S~3 we compare our predicted number of lensed sub-mm sources
to the observed number based on the 
SCUBA data. We follow the conventions that the Hubble
constant, $H_0$, is 100\,$h$\ km~s$^{-1}$~Mpc$^{-1}$, the present mean
density of the universe in units of the closure density is $\Omega_m$,
and the present normalized cosmological constant is $\Omega_\Lambda$.
In a flat universe, $\Omega_m+\Omega_\Lambda=1$.

\section{Expected Number of Lensed Sub-mm Sources}

In order to calculate the lensing rate for background sub-mm sources
due to foreground galaxy clusters, we model the
lensing clusters as singular isothermal spheres (SIS) and use the
analytical  filled-beam approximation (see, e.g., Fukugita et al.
1992). Our calculation is similar to that of C98 in which we calculated the
expected number of lensed optical arcs and radio sources
on the sky due to foreground galaxy clusters 
(see, also, Cooray, Quashnock \& Miller 1998).
We refer the reader to C98 for further details of our calculation, which
involves a description of the foreground galaxy clusters and their redshift
evolution using a PS analysis. The PS formalism utilizes normalizations
based on the local temperature function, and several scaling relations
between observed cluster properties such as mass and X-ray luminosity.

In order to describe the background sub-mm sources, we use the
redshift and number distribution observed towards the HDF sources 
by Hughes et al. (1998). 
The main advantage in using the HDF data is the availability of redshift
information for sub-mm sources. Also, HDF is one of the few areas where a deep
 survey at 850 $\mu$m down to a flux limit of $\sim$ 2 mJy 
has been carried out.
The HDF contains 5 sources with flux densities of the order $\sim$ 2 to 7
mJy. Hughes et al. (1998)  studied the probable redshifts of the detected
sources by considering the optical counterparts and assigning
probabilities for likely associations. We used the tabulated
redshifts in table 2 of Hughes et al. (1998) to describe the background
sources, and calculated the lensing probability using filled-beam formalism. 
Similar to C98, we
calculated the $<F(z_l)>$ parameter in lensing (see, C98) as a function
of the cosmological parameters 
by describing the foreground lensing clusters with a lower limit on the 
X-ray luminosity of $8 \times 10^{44}\,
h^{-2}\, {\rm ergs\, s^{-1}}$ in the EMSS bandpass of 0.3 to 3.5 keV.
This lower limit on the foreground
cluster X-ray luminosity allows a direct comparison between
the number  statistics of lensed sub-mm sources and that of the
luminous optical arcs.

We  calculated the expected number, $\bar N$,
of gravitationally lensed  radio sources on the sky as a function of
$\Omega_m$ and $\Omega_{\Lambda}$,
 and for a minimum amplification, $A_{\rm min}$, of 2
and 10 respectively. 
Since we are using the SIS model, the amplification is simply
equal to the ratio of length to width in observed lensing
arcs.
In Table 1, we list the expected number of  lensed sources on the sky
for $A_{\rm min}=2$ and 10.

\begin{table}
\caption[]{Predicted number of lensed sub-mm sources on the sky down to a flux density limit of 2 mJy at 850 $\mu$m. $\bar N_{\rm Planck}$ is the expected
 number of lensed sub-mm sources, with flux densities greater than 50 mJy
at 850 $\mu$m, towards clusters that are 
expected to be detected with Planck Surveyor (see, Section 3).}
\begin{tabular}{ccccc}
\noalign{\smallskip}
\hline
\noalign{\smallskip}
$\Omega_m$ & $\Omega_\Lambda$ & $\bar N(A_{\rm min} \geq 2)$ & $\bar N(A_{\rm min} \geq 10)$  & $\bar N_{\rm Planck}(A_{\rm min} \geq 2)$\\
\noalign{\smallskip}
\hline
\noalign{\smallskip}
0.1 & 0.0 & 14815 & 183 & 115 \\
0.2 & 0.0 & 9910 & 123 & 75 \\
0.3 & 0.0 & 6010 & 75 & 55 \\
0.4 & 0.0 &  3635 & 49 & 32 \\
0.5 & 0.0 &  2270 & 28 & 21 \\
0.6 & 0.0 & 1450 & 18 & 13 \\
0.7 & 0.0 & 790 & 10 & 9 \\
0.8 & 0.0 & 485 & 6 & 7 \\
0.9 & 0.0 & 310 &  4 & 4\\
1.0 & 0.0 & 225 & 3 & 2\\
0.1 & 0.9 & 270350 & 3340 &  2150  \\
0.2 & 0.8 & 101640 & 1255 & 785  \\
0.3 & 0.7 & 37835 & 468 & 290 \\
0.4 & 0.6 & 15050 & 186 & 105 \\
0.5 & 0.5 & 6470 & 80 & 45 \\
0.6 & 0.4 & 2975 & 37 & 25 \\
0.7 & 0.3 & 1440 & 18 & 15 \\
0.8 & 0.2 & 735 & 9 & 7 \\
0.9 & 0.1 & 390 & 5 & 4 \\
\noalign{\smallskip}
\hline
\end{tabular}
\end{table}

\section{Discussion}

Using the redshift and flux distribution observed for HDF sub-mm sources,
we have calculated the expected number of gravitationally lensed  sources
on the sky due to foreground clusters at a wavelength of 850 $\mu$m. 
By extrapolating the
observed properties towards the HDF to the whole sky, we have assumed 
that the HDF is a fair sample of the distant universe.
This assumption may be invalid given that the HDF was carefully selected
to avoid bright galaxies and radio sources. However, we have selected
to use the HDF data primarily because of the redshift
information for all detected sub-mm sources, which is currently not 
available for other  fields with sub-mm source observations.

We have predicted $\sim 1.5 \times 10^{4}$ lensed sub-mm sources
with flux densities greater than 2 mJy at 850 $\mu$m, and with 
amplifications greater than 2,
 on the sky towards clusters with X-ray luminosities greater than 
 $8 \times 10^{44}\, h^{-2}\, 
{\rm ergs\, s^{-1}}$, in the 0.3 to 3.5 keV EMSS band,
for a cosmology with $\Omega_m=0.4$ and $\Omega_\Lambda=0.6$.
This cosmology is consistent
with recent results based on lensing (CQM; Kochanek 1996), type Ia supernovae (Riess et al. 1998), and galaxy cluster baryonic fraction (Evrard 1997).  
The number with $A_{\rm min} > 4$ for the same cosmology is $\sim$ 1600,
while the number with $A_{\rm min} > 10$ is $\sim$ 200. 
The X-ray flux limit for foreground clusters
in our analysis is same as that of the clusters in the 
Le F\`evre et al. (1994) and Gioia \& Luppino (1994)
optical arc surveys, where 0.2 to 0.3 optical arc,
with length-to-width ratios greater than 10,  per cluster    
has been found down to a R band magnitude of $\sim$ 21.5.  
There are $\sim$ 7000 $\pm$ 1000 such clusters on the whole sky. 
We predict a lensing rate of
$\sim$ 2  sources per  cluster with amplifications greater than 2 down
to a flux limit of 2 mJy.

We compare our predicted number of lensed sources 
 to the observed number towards a sample
of galaxy clusters imaged with the SCUBA by Smail et al. (1997 \& 1998).
This sample contains 7 clusters with redshifts in the range $\sim$ 0.2
to 0.4. All of these clusters are well known lensing clusters in the
optical wavelengths. Unfortunately, this sample is incomplete either in
terms of X-ray luminosity or total mass. This incompleteness
doesn't allow us to perform a direct comparison between the
predicted and observed numbers. Out of the 7 clusters, 3 clusters have
X-ray luminosities greater than the lower limit imposed in our
calculation. Towards these three clusters, A370, A2390 \& A1835, there 
are 8 sub-mm sources, all of which may be gravitationally lensed.
This implies a total of $\sim 2 \times 10^{4}$ lensed sub-mm sources
on the whole sky.  Based on our lensing rate, we expect
$\sim$ 6 lensed sources towards
3 clusters; this exact number is strongly sensitive
to the cosmological parameters. Here, we have assumed a spatially-flat 
cosmological model with $\Omega_m=0.4$ and $\Omega_\Lambda=0.6$.
The predicted and observed numbers seem
 to be in agreement with each other
 for low $\Omega_m$ values in a flat universe
($\Omega_m+\Omega_\Lambda=1$). 

However, we cannot use the present observational
data to derive cosmological parameters for several reasons. These reasons
 include source contamination in the lensed source sample and systematic
biases in the foreground cluster sample. For example, it is likely that
the lensed source sample presented by Smail et al. (1998)
contain foreground and cluster-member sources. 
Since the foreground or cluster-member
sources are less bright than the background lensed sources, this
contamination is likely to be small (see, Blain 1997).
An additional systematic 
bias comes from the selection effects associated with the foreground 
 cluster sample. Since the observed clusters 
are  well known lensing clusters with high 
lensing rates at optical wavelengths,
 it is likely that there may be more lensed sub-mm 
sources towards these clusters than generally expected.
This increase is due to the complex potentials, dominated
by substructures, of these clusters.
Even though spherical models, such as the SIS model, 
can produce the overall lensed source statistics accurately, such
models cannot reproduce the observed arc statistics of clusters
with bimodal and other complex potentials.
Therefore, it is likely that the Smail et al. (1998) sample is biased towards
a higher number of lensed sub-mm sources, and by a simple
comparison based on observed and expected number of optical arcs for
these clusters,
we find that this overestimate can be as high as a factor of 2 to 3.

In order to constrain cosmological parameters based on statistics
of lensed sub-mm sources, results from a complete sample of
galaxy clusters, preferably from a large area survey,
 are needed. Further SCUBA observations of galaxy clusters,
perhaps the same cluster sample as the Le F\`evre et al. (1994) sample,
would be helpful in this regard. However, such a survey will
require a considerable amount of observing time, suggesting that current
instruments may not be able to obtain the necessary statistics.
However, in the near future there will be two opportunities to perform
 a large area sub-mm survey of galaxy clusters: 
the Planck Surveyor and the South Pole 10-m sub-mm telescope.

{\it South Pole 10 m sub-mm telescope}---The planned South Pole (SP) 
10-m sub-mm
telescope\footnote{http://cfa-www.harvard.edu/aas/tenmeter/tenmeter.html} 
is expected to begin observations around year 2003 (see, Stark et al.
1998). At 850 $\mu$m, it is expected that
within $\sim$ 90 hours  a square degree area will be surveyed
down to a flux limit of 1 mJy. 
Given the resolution and flux sensitivity, it is likely that
the SP telescope would be an ideal
instrument to survey either a sample of clusters or  random areas
to obtain lensed source statistics down to few mJy.
To obtain reliable values of the cosmological parameters based on
the sub-mm lensed source statistics, a survey of 
several hundred square degrees down to 
few $\times$ 1 mJy will be needed. A more direct approach within
a reasonable amount of observing time would be
to survey a carefully selected sample of galaxy clusters,
either based on X-ray luminosity or total mass, from which
lensed source statistics can easily be derived.

{\it Planck Surveyor}---Considering the amplification
distribution for SIS lens model, and the number counts defined
by Scott \& White (1998), we find that roughly 100 lensed
sub-mm sources  may be detected with the 
Planck Surveyor towards galaxy clusters (e.g., Bersanelli et al.
1996)\footnote{http://astro.estec.esa.nl/Planck/}. 
In Table 1, we list the number expected 
as a function of the cosmological parameters and
assuming that the Planck data will allow detection of sources
down to 50 mJy. However, given the
limited observational data on source counts at 850 $\mu$m, we note
that the predicted numbers may have large errors. We also note
that the Planck data will be highly confused, as the beam size
of Planck is $\sim$ few arcmins at 850 $\mu$m; even with $\sim$ 2 arcmin
physical pixels for high signal-to-noise  data, most of the sources
down to 50 mJy would be separated by only one or two pixels.
Assuming pixel sizes of the order beam size, the probability of finding
two sources with flux densities greater than 50 mJy in one Planck pixel would
 be $\sim$ 0.2 to 0.3. Thus, it is more likely that the Planck data will allow 
clear detection of sources down to $\sim$ 100 mJy, but with additional 
information, such as from other frequency channels and 
filtering techniques (see, e.g., Tegmark
\& de Oliviera-Costa 1998), it may be possible to lower this flux limit.

Also, it is likely that the lensed background sources
will contaminate the detection of 
Sunyaev-Zel'dovich (SZ) effect in galaxy clusters 
 (1998; see, Aghanim et al. 1997; Blain 1998). 
Since the SZ effect
and sub-mm sources have different spectral distributions, it is possible 
that with extra information on flux at different frequencies, both
the SZ effect and the source fluxes may be recovered.
However, due to the uncertainties involved with such an analysis,
as well as the effects due to source confusion,
it is likely that that Planck data would not readily 
allow an adequate determination of lensed sub-mm source statistics to 
constrain cosmological parameters.
It is more likely that the lensed sub-mm source catalog from Planck
 would be an important  tool to study the star-formation history at 
high redshifts; since lensing brightens sources, such a lensed source 
catalog will contain sub-mm sources  fainter than the current limit
predicted to be observable with Planck for unlensed sources.

\section{Summary}

Using the redshift and flux information for HDF sub-mm sources and
a Press-Schechter analysis for  foreground lenses,
we have calculated the expected number of lensed  sources at 850 $\mu$m 
 towards galaxy clusters.
In a cosmology with $\Omega_m=0.4$ and $\Omega_\Lambda=0.6$,
we predict $\sim 1.5 \times 10^{4}$ lensed sources towards clusters with
X-ray luminosities greater than $8 \times 10^{44}\, h^{-2}\,
{\rm ergs\, s^{-1}}$, and  with amplifications due to lensing
greater than 2. We have compared our predicted numbers to 
the observed number of lensed sub-mm sources towards
a sample of galaxy clusters. However, various biases in this observed sample
and possible source contamination, do not allow us to constrain cosmological
parameters based on current statistics.
We have briefly studied the possibility of using the Planck surveyor and the 
South Pole 10-m telescope data to
perform this task. A catalog of $\sim$ 100 lensed sources towards
clusters is likely to be a useful by-product of Planck.

\begin{acknowledgements}

I would like to thank Andrew Blain, the referee,
 for prompt refereeing of the paper. 
This study was  partially supported by
the McCormick Fellowship at the University of Chicago,
and a Grant-In-Aid of Research from the National Academy of 
Sciences, awarded through Sigma Xi, the Scientific Research Society.

\end{acknowledgements}

\end{document}